\documentclass{article}
\usepackage{jcappub}
\usepackage{subfig}
\usepackage{epsfig}
\usepackage{graphicx}
\usepackage{color}
\usepackage{amssymb}
\usepackage{amsmath}
\usepackage{mathtools}
\usepackage{hyperref}
\usepackage{url}
\usepackage[normalem]{ulem}
\usepackage{afterpage}
\usepackage{float}
\usepackage{textcomp}

\newcommand{\bn}{{\mathbf n}}

\newcommand{\HH}{{\cal H}}

\newcommand{\PP}{{\cal P}}
\newcommand{\RR}{{\cal R}}

\newcommand{\al}{\alpha}

\newcommand{\de}{\delta}

\newcommand{\La}{\Lambda}
\newcommand{\la}{\lambda}
\newcommand{\Om}{\Omega}

\newcommand{\si}{\sigma}

\newcommand{\be}{\begin{equation}}
\newcommand{\ee}{\end{equation}}

\newcommand{\bea}{\begin{eqnarray}}
\newcommand{\eea}{\end{eqnarray}}
\newcommand{\bean}{\begin{eqnarray*}}
\newcommand{\eean}{\end{eqnarray*}}

\newcommand*\braket[1]{\langle#1 \rangle}

\newcommand\spart{\;\raise1.0pt\hbox{$/$}\hskip-6pt\partial}
\definecolor{dred}{rgb}{0.8,0,0.1}
\definecolor{dgreen}{rgb}{0.0,0.5,0.1}

\definecolor{pdv}{rgb}{0.4,.7,0.6}

\title{ 
Perturbations of Cosmological Redshift Drift }

\author[1,2]{{Pedro} Bessa,}
\author[1]{Ruth Durrer and}
\author[1]{Dennis Stock}
\affiliation[1]{D\'epartement de Physique Th\'eorique and Center for Astroparticle Physics,\\
Universit\'e de Gen\`eve, 24 quai Ernest  Ansermet, 1211 Gen\`eve 4, Switzerland}
\affiliation[2]{PPGCosmo - UFES,
Universidade Federal do Espírito Santo,\\
Vitória, 29075-910, Brazil}

\emailAdd{Pedro.Bessa.1@etu.unige.ch}
\emailAdd{pedro.h.dutra@edu.ufes.br}
\emailAdd{ruth.durrer@unige.ch}
\emailAdd{dennis.stock@unige.ch}
\abstract{ In this paper we calculate the linear perturbations of the cosmological redshift drift. We show explicitly that our expressions are gauge-invariant and compute the power spectrum of the redshift drift perturbations and its correlations with galaxy number counts within linear perturbation theory. Our findings show that the perturbations are small, and that the peculiar velocity and acceleration terms are dominating  and cannot be neglected when modeling the full perturbative expression for the redshift drift. We also find that the cross-correlations with  galaxy number count fluctuations might increase the detectability of the effect and can help to separate the perturbative effects from the background cosmological redshift drift signal.}


\arxivnumber{}

\begin{document}
\maketitle
\flushbottom

\section{Introduction}\label{s:intro}
The redshift of a far away source increases as time goes on. Typically in about a Hubble time $1+z$ roughly doubles. Within short timescales of a few years this 'redshift drift' is very small. Nevertheless, this effect, originally pointed out by Sandage and McVittie \cite{Sandage:1962,McVittie:1962},  will probably be measurable in the near future~\cite{Liske:2008ph,Rocha:2022gog}. Near future experiments with extremely sensitive spectrography like the E-ELT or the SKAO, will
 reach the sensitivity to measure the redshift drift, e.g.  by monitoring Ly-$\al$ absorption lines of
distant quasars in the case of the E-ELT, within a time span of the order of a decade.

In a perfect Friedmann-Lema\^\i tre (FL) spacetime the redshift drift of a source at redshift $z$ is defined as $dz/d\tau_o$, where $\tau_o$ is the proper time of the observer. Denoting comoving time by $t$, using $dt_s = dt_o$  and $1+z = a_o/a_s$, so that $d\tau_s=a_sdt_s = (a_s/a_o)d\tau_o$, one obtains
\bea\label{e:dz-back}
\frac{dz}{d\tau_o} &=& \frac{d}{d\tau_o}\left(\frac{a_o}{a_s}\right) =\frac{1}{d\tau_o}\left[\frac{a_o(1+H_od\tau_o)}{a_s(1+H_s (a_s/a_o)d\tau_o)}-\frac{a_o}{a_s}\right] = H_o(1+z)-H_s \,.
\eea
Knowing the redshift drift of many sources
at different redshifts can therefore not only help us to measure the present Hubble parameter, $H_o\equiv H_0$, but especially also to determine its redshift dependence, $H(z)=H_s$. This is important, e.g. to constrain models of dark energy but also in homogeneous cosmological models, it can give us a more direct measurement of $H(z)$ which we presently mainly measure via Supernova data where we in principle have access to the luminosity distance which is proportional to an integral of $1/H(z)$.

In  general, the  angular dependence of the redshift drift can be used to test its isotropy and it is sensitive to large fluctuations on small scales in a statistically homogeneous and isotropic spacetime~\cite{Koksbang_2020_2,Koksbang_2021_prl}. In this work we study the redshift drift within first order perturbation theory, where backreaction from large perturbations on small scales are not considered.

 Cosmic structure changes the redshift via the induced velocities of source and observer and via the gravitational potential. To determine the precise expressions, we consider the redshift drift in a generic spacetime as given in~\cite{Heinesen_2021}, and apply it to a perturbed Friedmann universe.

Redshift drift in a perturbed FL universe has already been discussed before in~\cite{Marcori_2018}. But in this paper the authors derived an expression for the redshift drift valid for sources and observers with vanishing peculiar velocity in longitudinal gauge. We shall derive a fully gauge invariant expression and find that the velocity terms actually dominate the result by several orders of magnitude.

Our paper is structured as follows: in the next section, we introduce the redshift drift and give its general expression. In Section~\ref{sec:deltaz_perturbedFLRW} we calculate $dz/d\tau_o$ for a perturbed FL universe to first order in perturbation theory.
In Section~\ref{sec:power} we derive the power spectrum of the redshift drift and present numerical results using a modified version of {\sc class}. We also compute the cross-correlation between redshift drift and galaxy number counts. In Section~\ref{sec:con} we discuss our findings and conclude. 
\vspace{10pt}

\noindent{\bf Notation:} We work within a spatially flat FL spacetime with conformal time $t$ such that
$$ ds^2=a^2(t)[-dt^2+\de_{ij}dx^idx^j]\,.$$
The conformal Hubble parameter is $\HH =\dot a/a$, where an overdot denotes the derivative wrt.~conformal time $t$. Physical time is denoted by $\tau$. In the FL background, for a comoving observer,  $d\tau=a \,dt$ and the physical Hubble parameter is 
$$
H = \frac{1}{a}\frac{da}{d\tau} = \HH a^{-1} \,.
$$

\section{Redshift Drift}
\label{sec:Redshiftdrift}
\begin{figure}[h]
    \centering
    \includegraphics[scale=0.6]{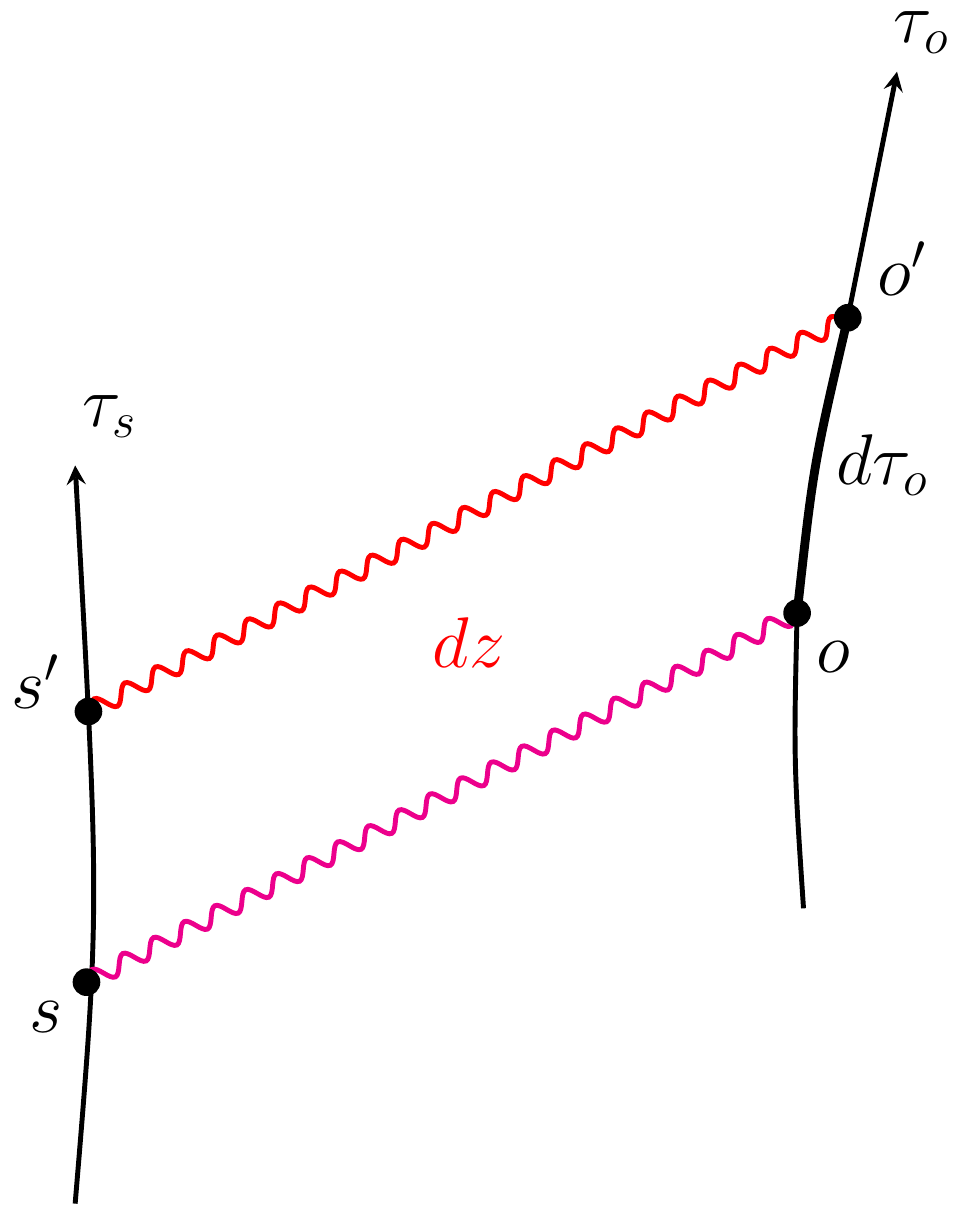}
    \caption{The change $dz$ of the redshift to a source $s$ per proper time interval $d\tau_o$ measured by the observer defines the redshift drift $\frac{dz}{d\tau_o}$. Note that in general the proper time of source and observer are different, $d\tau_o\neq d\tau_s$.}
    \label{fig:redshiftdrift}
\end{figure}

To determine the redshift drift in a generic spacetime, let us consider the worldline of a cosmic observer, in the following denoted by $o$, parametrised by its proper time $\tau_o$. The observer measures the redshift $z$ to a source, denoted by $s$. Recall that the redshift is  given in terms of the ratio of the observed and emitted photon energies, $E_o=-u\cdot k\big|_o$ and $E_s=-u\cdot k\big|_s$, where $\big|_o$ denotes evaluation at the observer of both 4-velocity $u^\mu$ and photon 4-momentum $k^\mu$, respectively at the source:
\begin{equation}
    1+z= \frac{E_s}{E_o}\,.
\end{equation}
The change of the redshift $dz$ per proper time interval of the observer $d\tau_o$ is called the redshift drift $\frac{dz}{d\tau_o}$, see  Fig.~\ref{fig:redshiftdrift}. 
Because redshifts are observable and the proper time $\tau_o$ is the time measured by the observer, the redshift drift is indeed a cosmological  observable. Note that in general there are several contributions to the redshift which vary with time, for example cosmic expansion, or peculiar velocities of observer and source. We will discuss all contributions in a perturbed FL universe  in detail in Section~\ref{sec:drift perturbed FLRW}. The redshift drift in a FL spacetime for an observer comoving with the fluid is derived in the introduction, Eq.~\eqref{e:dz-back}. 

Before computing the redshift drift in a linearly perturbed FL spacetime, we briefly review its general expression in an arbitrary spacetime in terms of kinematic quantities. This was originally derived in \cite{Heinesen:2020pms} and we shall use it  later. Let us denote the null vector generating the geodesic connecting source and observer along which the photon travels by $k$, and the (timelike) 4-velocities of source and observer by $u_s$ and $u_o$ respectively, which are normalised to one: $u\cdot u =-1$. The source and the observer move with the cosmic fluid to which we assign the 4-velocity field $u$. Along the photon geodesic we decompose $k$ into a part parallel to $u$ and a vector $e$, which is spacelike, orthogonal to $u$, and normalised to one, hence we have:
\begin{equation}
    k\cdot k=0\quad,\quad e\cdot e=1\quad,\quad u\cdot e=0\quad\text{with}\quad k=E(u-e)\quad,
    \label{eq:tetrad}
\end{equation}
where the  projection of $k$ onto the timelike direction $u$ is minus the photon energy measured by an observer moving with $u$, $E=-k\cdot u$. The redshift can then be expressed as
\begin{equation}
    1+z = \frac{E_s}{E_o}=\frac{(u\cdot k)|_s}{(u\cdot k)|_o}= \frac{d\tau_o}{d\tau_s}\quad,
    \label{eq:z2}
\end{equation}
where for the last equality we use that for photons $E=h\nu$, the frequency $\nu$ is the change of the phase $\varphi$ of the wave describing the photon with respect to proper time measured by observer, $\nu_{o}=\frac{d\varphi}{d\tau_{o}}$, respectively for the source, and that the phase is constant along a null geodesic. The redshift drift can then be written as
\begin{equation}
    \frac{dz}{d\tau_o} = \nabla_{u_o}\left(\frac{u\cdot k|_s}{u\cdot k|_o}\right)\quad.
    \label{eq:z}
\end{equation}
To carry out the derivative in (\ref{eq:z}) explicitly, we use the kinematic decomposition of $\nabla_\mu u_\nu$, see for example \cite{ellis_maartens_maccallum_2012,durrer_2020},
\begin{equation}
    \nabla_\mu u_\nu = \omega_{\mu\nu}+\Theta_{\mu\nu}-a_\mu u_\nu.
\end{equation}
This decomposition splits into an antisymmetric part $\omega_{\mu\nu}$ which is often referred to as vorticity, a symmetric part $\Theta_{\mu\nu}$, and a part proportional to the 4-acceleration $a^\mu=(\nabla_u u)^\mu$ quantifying the deviation of $u$ from geodesic motion. The symmetric part $\Theta_{\mu\nu}$ can be further decomposed into a trace $\theta$, the expansion scalar, and a traceless part $\sigma_{\mu\nu}$, referred to as shear tensor,
\begin{equation}
    \Theta_{\mu\nu} = \frac{1}{3}\theta h_{\mu\nu} +\sigma_{\mu\nu}\,,
\end{equation}
where $h_{\mu\nu}=g_{\mu\nu}+u_\mu u_\nu$ is the metric on the tangent space  normal to $u$. Eq.~(\ref{eq:z2}) implies that in full generality $\nabla_{u_o}=\frac{1}{1+z}\nabla_{u_s}$. Using also that $k$ follows a null geodesic $\nabla_k k=0$, we  arrive at the expression derived in \cite{Heinesen:2020pms},
\begin{align}
    \frac{dz}{d\tau_o} &= (1+z)\left[\mathfrak{h} -\frac{\nabla_e E}{E}\right]_o-\left[\mathfrak{h}-\frac{\nabla_e E}{E}\right]_s\,,~~\text{ with}
    \label{eq:exactdrift}\\
    \mathfrak{h}&:= \frac{1}{3}\theta+\sigma_{\mu\nu}e^\mu e^\nu -a_\mu e^\mu\,,
    \label{eq:generalised_h}
\end{align}
where $[\;]_{o/s}$ denotes that all terms inside the bracket are evaluated at the observer, respectively at the source and $\nabla_e$ is the covariant derivative in direction of the spacelike vector $e$. Note that in an FL spacetime $\mathfrak{h}=H$ and $\nabla_eE=0$ so that we recover \eqref{e:dz-back}. Expression~\eqref{eq:exactdrift} will be used to compute the redshift drift in perturbed FL in the next section.

\section{Redshift Drift in Perturbed FL}
\label{sec:deltaz_perturbedFLRW}
In this section, we first briefly introduce linear perturbations of the Friedmann metric and the energy-momentum tensor, before computing the redshift drift (\ref{eq:exactdrift}) for linear perturbations about an FL background. We also comment on how our results relate to previous findings in the literature.

\subsection{Perturbations of Metric and Energy-momentum Tensor}
We fix the background spacetime $\bar{g}_{\mu\nu}$ to be given by a FL spacetime, whose line element in conformal time $t$ reads
\begin{equation}
    d\bar{s}^2 = a^2(t)\left[-dt^2+\gamma_{ij}dx^i dx^j\right]\,,
\end{equation}
where $\gamma_{ij}$ is the spatial 3-metric of constant time hypersurfaces, which we consider to be spatially flat, i.e., $\gamma_{ij}=\delta_{ij}$. Here and in the following, overbars denote quantities evaluated in the background spacetime. Adding linear perturbations to the background in a general gauge, we obtain the metric 
\begin{equation}
\label{perturbed_metric}
   ds^2 = a^2(t)\left[-(1+2A)dt^2 -2B_idx^i dt + \left(\delta_{ij} + H_{ij} \right)dx^i dx^j\right]\,,
\end{equation}
see e.g. \cite{durrer_2020} for a more general and detailed account of cosmological linear perturbation theory. In addition to time, the perturbations also depend on position, because homogeneity and isotropy only hold for the background spacetime. In this work, we are interested in scalar perturbations and neglect vector and tensor perturbations. Vector perturbations decay in an FL background, see~\cite{durrer_2020} and the contribution from tensor perturbations is small compared to scalar perturbations, see e.g.~CMB data~\cite{Planck:2018jri}. Scalar spatial perturbations $H_{ij}$ can be decomposed into
\begin{equation}
    H_{ij}= H_L\delta_{ij} + H_{T\,ij}\qquad\text{with}\qquad H_{T\,ij}= \partial_i\partial_j H_T-\frac{1}{3}\delta_{ij} \Delta H_T\,.
\end{equation}
where $\Delta=\sum \partial_i^2$ denotes the flat space Laplacian.
Thus, scalar, linear metric perturbations are encoded in four functions: $A$, $B$ with $B_i=\partial_i B$, $H_L$ and $H_T$. In addition, the 4-velocity $u$ of the cosmic fluid is also perturbed,
\begin{equation}
\label{e:vel}
    \left( u^\mu\right) = \frac{1}{a}\left(1-A,v^i\right)\qquad\text{with}\qquad v_i=\partial_i v\,.
\end{equation}
The $u^0$ component is determined by the normalisation condition and the peculiar velocity is given by the potential $v$.

\subsection{Computation of the Redshift Drift in Perturbed FLRW}
\label{sec:drift perturbed FLRW}

Next, we perturb all quantities appearing in (\ref{eq:exactdrift}) to linear order and only at the end combine all terms into gauge invariant expressions. Because the final result is an observable, it must be gauge invariant. 

The perturbation of the 4-velocity field of source and observer is given in \eqref{e:vel}.
The perturbations of the photon 4-vector are determined by integrating the geodesic equation $\nabla_kk=0$ in the perturbed spacetime. One obtains
\begin{align}
    k^\mu &= \bar{k}^\mu +\delta k^\mu\qquad\text{with}\qquad \bar{k}^\mu = \frac{1}{a^2}(1,\bar{n}^i)\qquad\text{and}\\
    \delta k^0 &= \frac{1}{a^2}\delta n^0 = \frac{1}{a^2}\left\{ \left[ -2A-B_j\Bar{n}^j \right]_s^o-\int_s^o d\lambda\; \left[\dot{H}_L-\dot{A}+\dot{H}_{T ab}\Bar{n}^a\Bar{n}^b-\dot{B}_a\Bar{n}^a\right] \right\}\,,\label{eq:deltak0}\\
    \delta k^i &= \frac{1}{a^2}\delta n^i =\frac{1}{a^2}\left\{ \left[-B^i+2H_L\Bar{n}^i+H_T^{ij}\Bar{n}_j \right]_s^o-\int_s^o d\lambda\; \left[\partial^i(H_L-A) +\Bar{n}^a\Bar{n}^b\partial^i H_{T ab}\right] \right\} \,,
\end{align}
where $\bar{n}^i$ is the unperturbed direction of the photon geodesic satisfying $\Bar{n}_i\bar{n}^i=1$. The integrals are performed along the unperturbed photon geodesic in the background spacetime from source to observer, and $\lambda$ denotes the affine parameter along the null geodesic. The spacelike vector $e$ orthogonal to $u$ from (\ref{eq:tetrad}) is then given by
\begin{equation}
    e^\mu= u^\mu - \frac{1}{E}k^\mu = \frac{1}{a}\left(\bar{n}^i(B_i-v_i)\;,\; -\bar{n}^i-\delta n^i+v^i+\bar{n}^i\left(A+\bar{n}^j(B_j-v_j)+\delta n^0\right)\right)\,,
\end{equation}
where we used the first order expression for the photon energy
\begin{equation}
    E=-k_\mu u^\mu = \frac{1}{a}\left[ 1+A+\bar{n}^i(B_i-v_i)+\delta n^0 \right]\,.
\end{equation}
With that, we can determine the remaining quantities in (\ref{eq:exactdrift}) to linear order,
\begin{align}
    a^\mu &= \left(0\;,\; -\frac{1}{a^2}\partial^i\left[\dot{v}-\dot{B}-A+\mathcal{H}(-B+v)\right]\right)\\
    \theta &= \frac{1}{a}\left[3\mathcal{H}(1-A)-\Delta v+3\dot{H}_L \right]\\
    \sigma_{00} &= \sigma_{0i}=\sigma_{i0} = 0\\
    \sigma_{ij} &= a\left(\partial_i\partial_j-\delta_{ij}\frac{1}{3}\Delta \right)(\dot{H}_T-v)\\
    \nabla_e E &= -\frac{1}{a^2}\left\{ \bar{n}^i\partial_i(A+\delta n^0)+\bar{n}^i\bar{n}^j\partial_i(B_j-v_j)+\mathcal{H}\bar{n}^i(B_i-v_i) \right\}\,.
\end{align}
 While it is easy to verify that $a^\mu$, $\sigma_{\mu\nu}$ and $\nabla_e E$ are gauge invariant (as they should be according the the Stewart--Walker lemma, since they vanish in the Friedmann background),  $\theta$ is not. Therefore $\mathfrak{h}$ is not gauge invariant.
Putting together the individual terms to obtain the generalised Hubble function $\mathfrak{h}$ (\ref{eq:generalised_h}), and using the relations given in Appendix \ref{ap:perturbationvariables} to convert the expression to gauge-invariant perturbation variables whenever possible, we find
\begin{equation}
    \mathfrak{h}(t) = \frac{1}{a}\left[ \mathcal{H}(1-A) +\dot{\mathcal{R}}-\bar{n}^i\partial_i\left(\dot{V}+\mathcal{H}V-\Psi\right)-\bar{n}^i\bar{n}^j\partial_i\partial_j V \right]\quad,
    \label{eq:h}
\end{equation}
with the gauge-invariant velocity potential $V$, the spatial curvature $\mathcal{R}$ of the spacelike hypersurface of constant time, and the Bardeen potentials $\Phi$, $\Psi$. Note that $A$ and $\mathcal{R}$ are still gauge dependent, hence \eqref{eq:h} is not gauge invariant as expected. 

In a next step, we express $\nabla_e E$ in terms of gauge-invariant variables. In order to achieve that, we first express the integrand of (\ref{eq:deltak0}) in terms of the Bardeen potentials using the relations given in Appendix \ref{ap:perturbationvariables}:
\begin{equation}
    \dot{H}_L-\dot{A}+\dot{H}_{T ab}\Bar{n}^a\Bar{n}^b-\dot{B}_a\Bar{n}^a = -\dot{\Phi}-\dot{\Psi}-\dddot{H}_T+\ddot{B}+\bar{n}^i\bar{n}^j\partial_i\partial_j \dot{H}_T+\bar{n}^i\partial_i \dot{B}\quad.
\end{equation}
Next, we use the partial integration formula for a first order quantity $f$
\begin{equation}
    \frac{df}{d\lambda}=\nabla_n f= \dot{f}+\bar{n}^i\partial_i f\quad\Rightarrow\quad \int_s^o d\lambda\, \bar{n}^i\partial_i f = f_o-f_s-\int_s^o d\lambda\, \dot{f}\quad,
    \label{eq:partint}
\end{equation}
in order to turn the spatial derivatives in  $\bar{n}^i\partial_i \dot{B}$ and $\bar{n}^i\bar{n}^j\partial_i\partial_j \dot{H}_T$ into time derivatives, yielding
\begin{equation}
    \int_s^o d\lambda\; \left[\dot{H}_L-\dot{A}+\dot{H}_{T ab}\Bar{n}^a\Bar{n}^b-\dot{B}_a\Bar{n}^a\right]= -\int_s^o d\lambda\, (\dot{\Phi}+\dot{\Psi})+(\dot{B}_o-\dot{B}_s)+\bar{n}^i\partial_i(\dot{H}_T^o-\dot{H}_T^s)-(\ddot{H}_T^o-\ddot{H}_T^s)\quad.
\end{equation}
In a next step, we evaluate
\begin{equation}
    \delta n^0|_s^o = \left[ -2A-B_j\Bar{n}^j \right]_s^o-\int_s^o d\lambda\; \left[\dot{H}_L-\dot{A}+\dot{H}_{T ab}\Bar{n}^a\Bar{n}^b-\dot{B}_a\Bar{n}^a\right]\,.
\end{equation}
In what follows we neglect the terms evaluated at the observer, because they give rise to an unmeasurable monopole, and to a  dipole contribution due to the observer velocity. This dipole is interesting as it provides an alternative measure of the observer velocity. It has been studied in~\cite{Inoue:2019qvy}.
Neglecting the observer terms in $\de n^0$ we obtain
\begin{equation}
    \delta n^0= -2A -\bar{n}^i\partial_i \sigma_t +\dot{\sigma}_t - \int_s^o d\lambda\,(\dot{\Phi}+\dot{\Psi})\quad,
\end{equation}
where we dropped the source label $s$, such that if not specified differently, all terms are to be evaluated at the source. We have also introduced the shear of the constant time hypersurface $\si_t$ which is not gauge invariant, see Appendix \ref{ap:perturbationvariables}. Using again the relations given in Appendix \ref{ap:perturbationvariables}, we arrive at the following expression for $\nabla_e E$ in terms of gauge-invariant variables:
\begin{equation}
    \nabla_e E = \frac{1}{a^2}\bar{n}^i\partial_i\left[\Psi-\mathcal{H}V-\bar{n}^j\partial_j V+\int_s^o d\lambda\,(\dot{\Phi}+\dot{\Psi})\right]\quad.
    \label{eq:nablaE}
\end{equation}
Not surprisingly, $\nabla_e E$ which vanishes on the background is a gauge invariant quantity (a consequence of the Stewart Walker lemma).

However, $\mathfrak{h}$ given above depends on the chosen time-slicing. 
So far, we evaluated all quantities at fixed time $t$ and as such evaluated them on hypersurfaces of constant time. However, constant time hypersurfaces are observationally not accessible, instead, we can observe hypersurfaces of constant observed redshift $z$. Since  time is related to the background redshift, $t(\bar{z})$, and $z=\bar{z}+\delta z$, we have to convert a quantity $f(t)$ evaluated on constant time hypersurfaces via
\begin{equation}
    f(t)\rightarrow f(t(\bar{z})) \equiv f(\bar{z}) = f(z-\delta z)\approx f(z) - \frac{d\bar{f}}{d\bar{z}}\,\delta z
    \label{eq:conversion}
\end{equation}
to first order. Note that within linear perturbation theory the redshift correction term proportional to $\delta z$ is only relevant for background quantities, because for first order terms, the correction would be of 
second order and therefore it can be neglected.  As $\nabla_e E$  in our expression (\ref{eq:exactdrift}) for the redshift drift is already first order, its conversion gives no first order contribution and only $\mathfrak{h}$ has to be converted. According to (\ref{eq:conversion})
\begin{equation}
    \mathfrak{h}(t)\approx \mathfrak{h}(z)-\frac{d\bar{\mathfrak{h}}}{d\bar{z}}\,\delta z = \mathfrak{h}(z)+\left(\frac{\dot{\mathcal{H}}}{\mathcal{H}}-\mathcal{H}\right)\,\delta z\quad.
\end{equation}
The redshift fluctuation $\delta z$ was already computed e.g. in \cite{Bonvin_2011} and reads, neglecting again observer monopole and dipole terms, 
\begin{equation}
    \delta z = -(1+z)\left[H_L+\frac{1}{3}H_T+n^i\partial_i V+\Phi+\Psi+\int_s^o d\lambda\,(\dot{\Phi}+\dot{\Psi}) \right]\,.
\end{equation}
Using the relations in Appendix \ref{ap:perturbationvariables}, $\mathfrak{h}(z)$ can be expressed in terms of gauge-invariant variables,
\begin{align}
    \mathfrak{h}(z) =& \frac{1}{a}\left\{ \mathcal{H}-\dot{\Phi}-\bar{n}^i\partial_i\left(\dot{V}+2\mathcal{H}V-\Psi\right)-\bar{n}^i\bar{n}^j\partial_i\partial_j V-\frac{\dot{\mathcal{H}}}{\mathcal{H}}\left(\Psi-\bar{n}^i\partial_i V\right)+ \right.\nonumber \\
    & \left. \qquad + \left(\mathcal{H}-\frac{\dot{\mathcal{H}}}{\mathcal{H}}\right)\int_s^o d\lambda\, (\dot{\Phi}+\dot{\Psi}) \right\}\,,
\end{align}
where again observer monopole and dipole terms are neglected. Finally, putting all contributions to (\ref{eq:exactdrift}) together and neglecting observer monopole and dipole terms, we arrive at the following, manifestly gauge-invariant expression for the redshift drift as function of the observed redshift:

\begin{equation}
    \frac{dz}{d\tau_o} =(1+z)\left[-\mathcal{H}+\Dot{\Phi}+\bar{n}^i\partial_i(\Dot{V}+\mathcal{H}V) +\frac{\Dot{\mathcal{H}}}{\mathcal{H}}(\Psi-\bar{n}^i\partial_iV)+
	 \left(\frac{\Dot{\mathcal{H}}-\mathcal{H}^2}{\mathcal{H}}+\bar{n}^i\partial_i \right)\int_{s}^{o}\hspace{-1.3mm}d\lambda\left(\Dot{\Phi}+\Dot{\Psi}\right) \right]\,.
\label{eq:main_redshiftdrift}
\end{equation}
Note that the zeroth order expression of this no longer coincides with \eqref{e:dz-back}, both expressions differ by an observer monopole term, but they agree at zeroth order after this observer term is dropped. We then get $-H_s = -(1+z)\HH$.
Here $\frac{dz}{d\tau_o}$ is considered as a function the observed redshift $z$ and the observed incoming direction of the photon $\bn$.

It is worth commenting on the individual contributions. As already mentioned, the leading order contribution is  given by the background Hubble function $-H_s=-(1+z)\mathcal{H}$ due to the isotropic and homogeneous expansion of the Universe, compare with (\ref{e:dz-back}) neglecting the observer term. The contributions proportional to $\bar{n}^i\partial_i V$ are Doppler terms arising from the source's peculiar velocity which affects the redshift, and $\bar{n}^i\partial_i \dot V$ is a peculiar acceleration term. The terms involving the Bardeen potentials $\Phi$, $\Psi$ account for local fluctuations in the gravitational potential at the source position, whereas $\int_{s}^{o}d\lambda\left(\Dot{\Phi}+\Dot{\Psi}\right)$ is the integrated Sachs-Wolfe effect accounting for the time variations of the Bardeen potentials integrated along the line of sight.
In fact, we will use a slightly different expression in the next section, where the spatial derivative of the integrated Sachs-Wolfe effect is rewritten via the partial integration formula (\ref{eq:partint}),
\begin{equation}
    \bar{n}^i\partial_i\left[\int_s^o d\lambda\, (\dot{\Phi}+\dot{\Psi})\right] = -(\dot{\Phi}+\dot{\Psi}) -\int_s^o d\lambda\, (\ddot{\Phi}+\ddot{\Psi})\,.
\end{equation}
Observer monopole terms are again neglected. This brings (\ref{eq:main_redshiftdrift}) to the form
\begin{align}
    \frac{dz}{d\tau_o} =(1+z)\bigg[&-\mathcal{H}-\Dot{\Psi}+\bar{n}^i\partial_i(\Dot{V}+\mathcal{H}V) +\frac{\Dot{\mathcal{H}}}{\mathcal{H}}(\Psi-\bar{n}^i\partial_iV)\nonumber\\
    &+\frac{\Dot{\mathcal{H}}-\mathcal{H}^2}{\mathcal{H}}\int_{\mathcal{S}}^{\mathcal{O}}d\lambda\left(\Dot{\Phi}+\Dot{\Psi}\right)-\int_s^od\lambda\,(\ddot{\Phi}+\ddot{\Psi}) \bigg]\,.
    \label{eq:reddrift_final}
\end{align}
Even though this formula is a bit more cumbersome than (\ref{eq:main_redshiftdrift}), it is useful as in the numerical analysis we shall neglect time derivatives of the Bardeen potentials which are known to be very small.

There are previous results on the redshift drift in the literature, most notably in \cite{Marcori_2018}, where the authors compute the redshift drift in a perturbed Friedmann spacetime on constant time hypersurfaces, neglecting peculiar velocities. It turns out that if we disregard their monopole and dipole terms at the observer, we match their result when converting it from constant conformal time hypersurfaces to constant redshift hypersurfaces. After neglecting our velocity terms ($V=0$) in (\ref{eq:main_redshiftdrift}), both results agree. As it is shown in the next section, however, the contributions coming from the source velocity  dominate the redshift drift perturbations and therefore they have to be included.

\section{Angular Power Spectra}\label{sec:power}
Next, we study the angular power spectrum of redshift drift perturbations. In fact, we are interested in the spectra of the relative fluctuations with respect to the background value,
\begin{align}
    \delta\dot{z}&:=\frac{\frac{dz}{d\tau_o}-\frac{d\bar z}{d\bar\tau_o}}{\frac{d\bar z}{d\bar\tau_o}} \nonumber\\
    &= -\frac{1}{\mathcal{H}}\Bigg[-\Dot{\Psi}+\bar{n}^i\partial_i \Dot{V} +\frac{\Dot{\mathcal{H}}}{\mathcal{H}}\Psi
    +\left(\frac{\Dot{\mathcal{H}}-\mathcal{H}^2}{\mathcal{H}}\right)\left(\int_{\mathcal{S}}^{\mathcal{O}}d\lambda\left(\Dot{\Phi}+\Dot{\Psi}\right) -\bar{n}^i\partial_i V\right)-\int_s^od\lambda\,(\ddot{\Phi}+\ddot{\Psi})\Bigg]\;,
    \label{eq:zdot}
\end{align}
where $\frac{d z}{d\tau_o}$ is the redshift drift in the perturbed spacetime (\ref{eq:reddrift_final}), and $\frac{d\bar z}{d\bar\tau_o}=-(1+z)\mathcal{H}$ the redshift drift in the background spacetime, neglecting the observer term, $H_0(1+z)$.

Because $\delta\dot{z}(\mathbf{n},z)$ is a function of  directions on the sphere, $\mathbf{n}$, we can decompose it into spherical harmonics, $Y_{\ell m}(\mathbf{n})$,
\begin{equation}
\label{expansion_sph_har}
    \delta\dot{z}(\mathbf{n},z) = \sum_{\ell,m}^\infty a_{\ell m}^{\delta\dot{z}}(z)Y_{\ell m}(\mathbf{n})\,, 
\end{equation}
where the expansion coefficients $a_{\ell m}^{\delta\dot{z}}$ are given by
\begin{equation}
\label{alm}
    a_{\ell m}^{\delta\dot{z}}(z) = \int d\Omega_\mathbf{n}\, \delta\dot{z}(\mathbf{n},z)Y^*_{\ell m}(\mathbf{n})\,.
\end{equation}

The angular power spectrum $C_\ell(z)$ is then defined as
\begin{equation}
    C_\ell^{\delta\dot{z}}(z) := \langle|a_{\ell m}^{\delta\dot{z}}|^2\rangle\,.
\end{equation}
Here $\braket{\,}$ denotes the average over a statistical ensemble. For statistically homogeneous and isotropic distributions, these expectation values are independent of $m$, see e.g.~\cite{durrer_2020}.

Using the Fourier decomposition of a general function $X$ of the spacetime coordinates,
\begin{equation}
    X(\mathbf{x},t) = \frac{1}{(2\pi)^3}\int X(\mathbf{k},t)e^{-i\mathbf{k}\cdot\mathbf{x}}d^3k\quad,
\end{equation}
we can rewrite the expansion coefficients $a_{\ell m}$ of the terms appearing in (\ref{eq:zdot}), using (\ref{alm}), as
\begin{align}
\label{coeff}
    a_{\ell m}^\Psi(z_s) &= \frac{i^\ell}{2\pi^2}\int d^3k j_\ell(kr_s)\Psi(\mathbf{k},t_s)Y^*_{\ell m}(\mathbf{\Hat{k}}),\qquad a_{\ell m}^{\Dot{\Psi}}(z_s) &= \frac{i^\ell}{2\pi^2}\int d^3k j_\ell(kr_s)\Dot{\Psi}(\mathbf{k},t_s)Y^*_{\ell m}(\mathbf{\Hat{k}}), \notag\\
    a_{\ell m}^{\mathbf{V}\cdot\mathbf{n}}(z_s) &= \frac{i^\ell}{2\pi^2}\int d^3k j'_\ell(kr_s)V(\mathbf{k},t_s)Y^*_{\ell m}(\mathbf{\Hat{k}}),\qquad a_{\ell m}^{\Dot{\mathbf{V}}\cdot\mathbf{n}}(z_s) &= \frac{i^\ell}{2\pi^2}\int d^3k j'_\ell(kr_s)\Dot{V}(\mathbf{k},t_s)Y^*_{\ell m}(\mathbf{\Hat{k}}),\notag\\
\end{align}
where $\mathbf{V}\cdot\mathbf{n} = n^i\partial_i V$, and $'$ denotes derivation with respect to the argument.

Next, we assume that each of these fluctuations at redshift $z$ is related to the initial primordial  perturbation $\Psi_\mathrm{in}$ via a respective transfer funtion $T(k,z)$, which propagates the perturbation up to the desired redshift and therefore contains information about the evolution history of the Universe,
\begin{align}
\label{transfer_func_psi}
    \Psi(\mathbf{k},z) &=  T_\Psi(k,z)\Psi_{\text{in}}(\mathbf{k})\\
    \label{transfer_func_phi}
    \Phi(\mathbf{k},z) &=  T_\Phi(k,z)\Psi_{\text{in}}(\mathbf{k})\\
    \label{transfer_func_v}
    V(\mathbf{k},z) &=  T_V(k,z)\Psi_{\text{in}}(\mathbf{k}).
\end{align}
This is certainly the case for adiabatic perturbations of standard single field slow roll inflation.
In this case the initial power spectrum is characterized by a simple power law,
\begin{equation}
\label{primordial_ps}
    k^3\langle\Psi_{\text{in}}(\mathbf{k})\Psi^*_{\text{in}}(\mathbf{k})\rangle = k^3P_\Psi\equiv(2\pi)^3\delta(\mathbf{k}-\mathbf{k'})A\left(\frac{k}{k_*}\right)^{n_S-1}\,,
\end{equation}
with amplitude $A$ at a fixed pivot scale $k_*$ and a spectral index $n_S$.

In the actual redshift drift perturbation power spectrum the amplitude $A$ is the one of the dimensionful initial power spectrum of the Bardeen potential \eqref{primordial_ps} $P_\Psi$, which in a $\La$CDM universe is related to the dimensionless curvature power spectrum via
\begin{align}
k^3P_\Psi &= 2\pi^2\left(\frac{3}{5}\right)^2\PP_\RR\\
A\left(\frac{k}{k_*}\right)^{n_S-1} &= 2\pi^2\left(\frac{3}{5}\right)^2A_S\left(\frac{k}{k_*}\right)^{n_S-1} \,,
\end{align}
where $A_S$ is the amplitude of the curvature power spectrum and $n_S$ is the spectral index. Their values have been measured from CMB fluctuations, see~\cite{Planck_18}.

The transfer functions $T_\Psi$, $T_\Phi$, and $T_V$ are related via the first order Einstein equations, and for the standard $\La$CDM universe with only one degree of freedom in the perturbations, they are given, e.g., in terms of  the transfer function of the Bardeen potential $T_\Psi$,~\cite{Bonvin_2011}:
\begin{subequations}
\begin{align}
    T_\Phi&=  T_\Psi\quad,\label{eq:T_phi}\\
    T_V&= \frac{2a}{3\Omega_m}\frac{k}{\mathcal{H}_0^2}\left(\mathcal{H}T_\Psi +\Dot{T}_\Psi   \right)\label{eq:T_v}\\
    kT_\Psi&=\dot T_V +\HH T_V \label{pec_acc}.
\end{align}
\end{subequations}
In terms of the  time-independent transfer function $T(k)$, we can write the  transfer function for the Bardeen potential $T_\Psi$ as \cite{dodelson2003,Bonvin_2011}

\begin{equation}\label{e:TPsi}
    T_\Psi(k,z) = \frac{9}{10}\frac{D_+(a)}{a}T(k)\quad,
\end{equation}
where $D_+(a)$ is the growth rate for the given cosmology and $z=a_0/a-1$.

The transfer functions in \eqref{transfer_func_psi}-\eqref{transfer_func_v} can be obtained in multiple ways: either numerically through Boltzmann codes such as {\sc{class}} \cite{CLASS2}, assuming an analytical ansatz for the matter transfer function $T(k)$ such as the Eisenstein-Hu formula \cite{Eisenstein_1998}, or using one of the more recent algebraic expressions obtained through so called  genetic algorithms \cite{Orjuela22}.  We have checked that for our cosmological parameters the Eisenstein-Hu (EH) transfer function agrees within less than one percent with the numerical transfer function derived from the {\sc{CLASS}} code, which is precise enough for the numerical results presented in this paper. However, for our final numerical plots  we chose to modify the {\sc{CLASS}} transfer function module originally called {\sc{CLASS}}gal and described in detail in~\cite{DiDio:2013bqa}.

This allows us to write the power spectrum of the redshift drift fluctuation \eqref{eq:zdot} at redshift $z_s$ of the source, $C_\ell(z_s)$  as 

\begin{equation}
    C_\ell(z_s) = \frac{2A}{\pi}\int \frac{dk}{k}\left(\frac{k}{k_*}\right)^{n_S-1}|F_\ell(k,z_s)|^2,
    \label{power_spectra_rd}
\end{equation}

where
\begin{align}
    F_\ell(k,z_s) = -\frac{1}{\mathcal{H}}\Bigg[&j_\ell(kr_s)\left( -\dot{T}_\Psi+\frac{\dot{\mathcal{H}}}{\mathcal{H}}T_\Psi\right)+ j'_\ell(kr_s)\left( \dot{T}_V-\frac{\Dot{\mathcal{H}}-\mathcal{H}^2}{\mathcal{H}}T_V\right)\nonumber\\
    &+\frac{\Dot{\mathcal{H}}-\mathcal{H}^2}{\mathcal{H}} \int_s^o d\lambda\, j_\ell(kr_s) \left( \dot{T}_\Phi+\dot{T}_\Psi\right) -\int_s^o d\lambda\, j_\ell(kr_s)\left( \ddot{T}_\Phi+\ddot{T}_\Psi\right)\Bigg]\quad.
    \label{eq:F_ell}
\end{align}
\

In the following paragraphs, we use a modified version of {\sc{CLASS}} in order to study the dependence of the redshift drift on redshift and angular scales, as well as its cross correlations with galaxy number count fluctuations. To that end, we apply a Gaussian window function in redshift of width $\Delta z=0.01$ in all figures below. The numerical values for the cosmological background parameters underlying this analysis are based on the latest Planck results and are summarised in Table~\ref{tab:parameters}, in particular the amplitude $A_S$ and spectral index $n_S$ of the primordial power spectrum, as well as the Planck values for $h$, $z_\text{eq}$, $k_\text{eq}$, $\Om_\text{b}$ and $\Om_\text{cdm}$ which are needed for the transfer functions.

\begin{table}
\begin{center}
\begin{tabular}{|l|l|l|}
\hline
Parameter& Value & Units\\
\hline
$\Omega_b$&$0.022383$&$h^{-2}$\\
$\Omega_{\text{CDM}}$&$0.12011$&$h^{-2}$\\
$h$&$0.7$&$10^{-2}{\rm km/s/Mpc}$\\
$A_S$&$2.100549\cdot10^{-9}$& - \\
$n_S$&$0.9660499$& - \\
$z_{\text{eq}}$&$3387$& - \\
$k_{\text{eq}}$&$0.010339$& $h/{\rm Mpc}$\\
\hline
\end{tabular}
\end{center}
\caption{Cosmological parameters underlying the numerical results in Section~\ref{sec:power}.}
\label{tab:parameters}
\end{table}

\subsection{Power spectra at a single redshift and correlated at diffent redshifts}

\begin{figure}[hb]
    \centering
    \includegraphics[scale=.7]{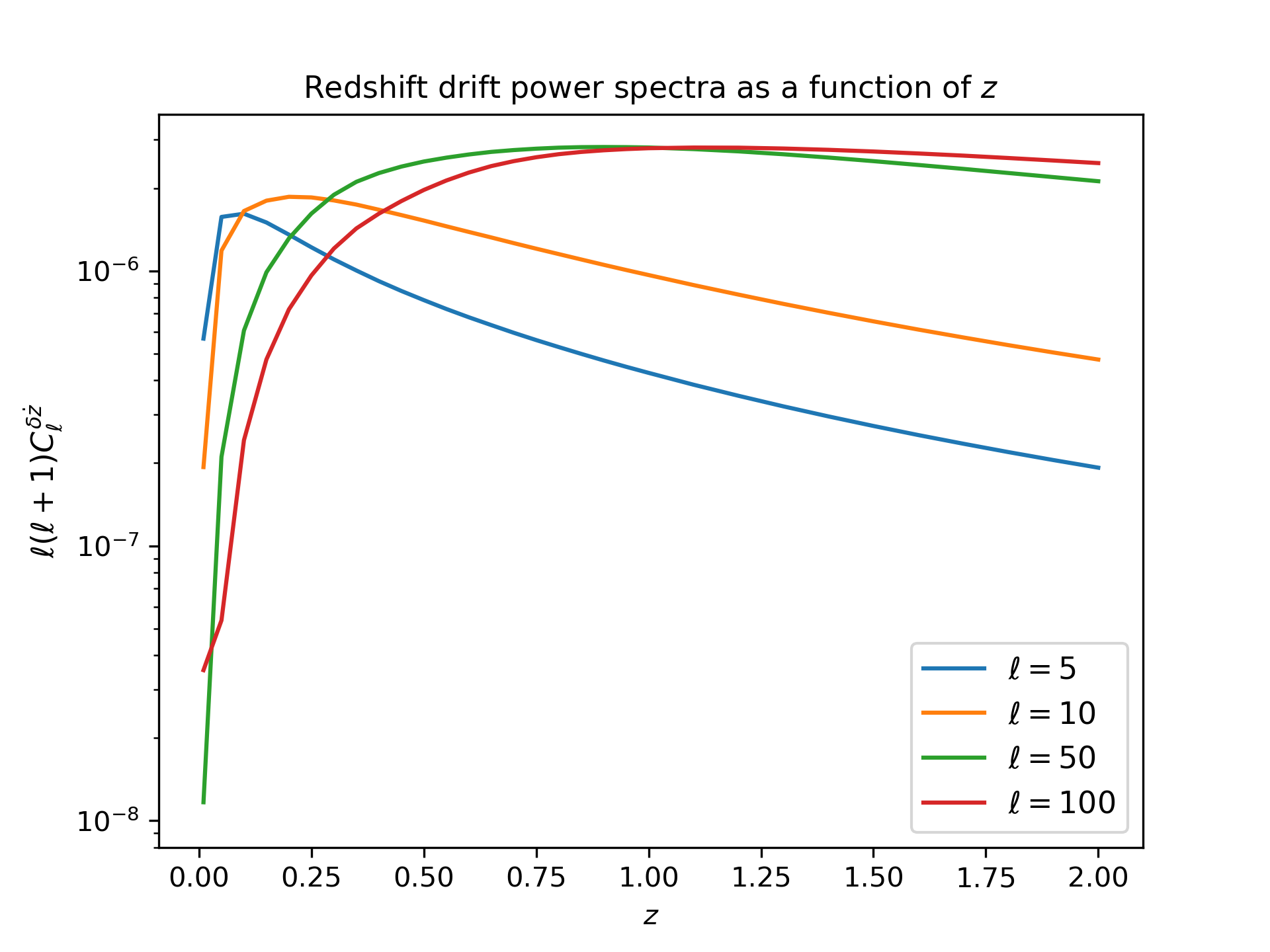}
    \caption{Evolution of the perturbed redshift drift  power spectra $C_\ell^{\delta\dot z}(z)$ as a function of redshift, for different multipoles $\ell$. }
    \label{redshift_drift_z}
\end{figure}

In Fig.~\ref{redshift_drift_z} we plot the angular power spectrum as a function of redshift for some fixed multipoles, $\ell$, neglecting the integrated terms in the second line of \eqref{eq:F_ell}. We have checked that the integrated terms $\Dot{T}_\Psi$ and $\Ddot{T}_\Psi$ in \eqref{eq:main_redshiftdrift} contribute less than $1\%$  to the total power spectrum over the $\ell$ range of $2-300$, so we neglect them  in the numerical evaluations. At redshifts $z\gtrsim 1$, the universe is matter dominated and the Bardeen potentials in \eqref{power_spectra_rd} are nearly constant~\cite{durrer_2020}. For low multipoles, $\ell\leq 10$, after an initial increase up to $z\simeq 0.05$, the redshift drift decays rapidly. For the higher multipoles, $\ell\geq 50$, after an initial growth the redshift drift nearly levels off. 

\begin{figure}[ht]
    \centering
    \includegraphics[scale=0.8]{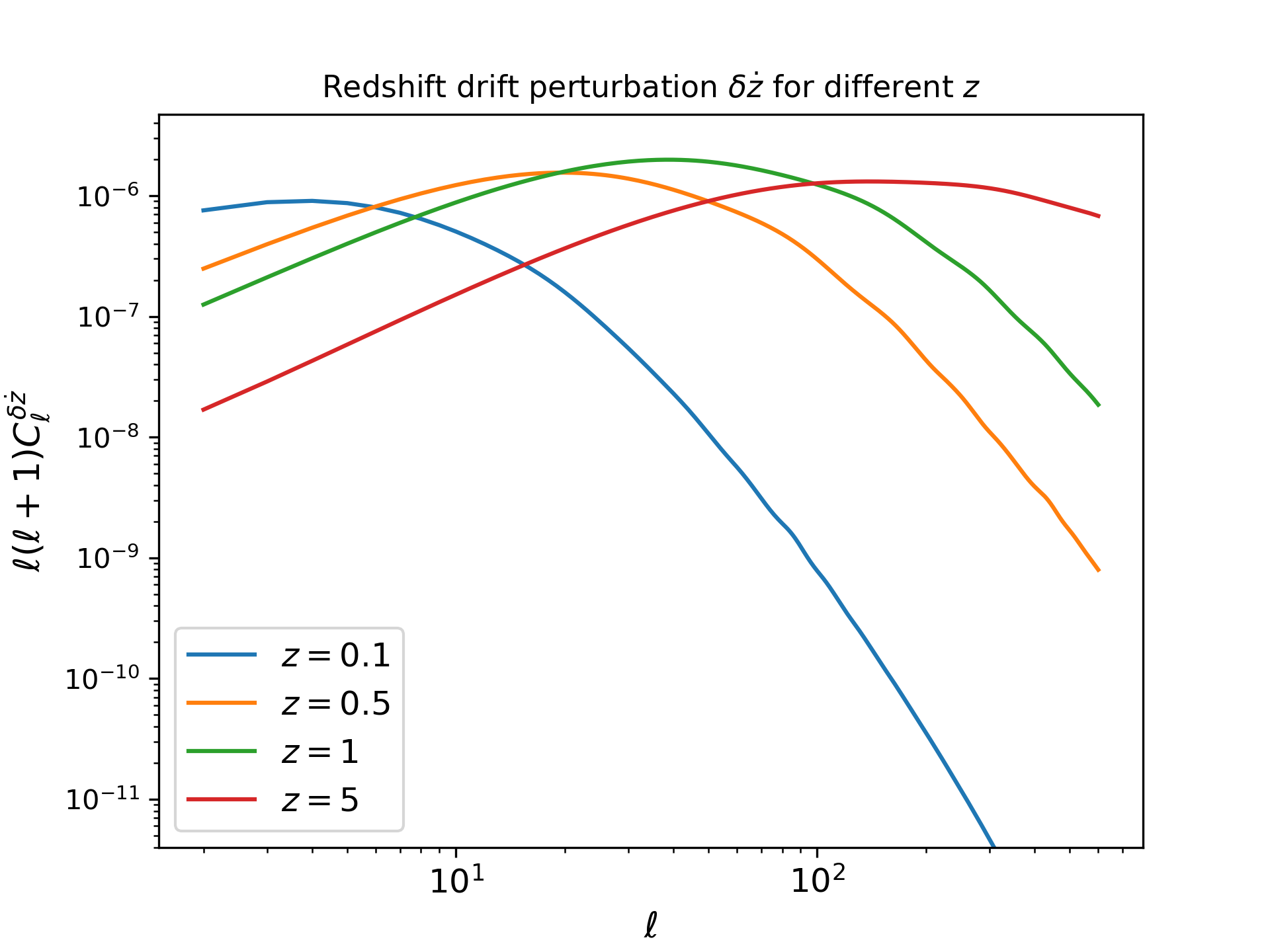}
    \caption{Redshift drift perturbation power spectra for a given set of redshifts $z_s$ and scales $\ell\leq 600$.}
    \label{power_spectra_l_500}
\end{figure}

We plot the full spectra for a set of redshifts and $\ell\leq 500$ in Fig.~\ref{power_spectra_l_500}.
Clearly, the perturbations are small, with amplitudes less than $10^{-3}$ on all scales. While the low redshift fluctuations dominate on large angular scales $\ell\lesssim 30$, the higher redshift spectra $z\geq 1$ peak at somewhat smaller angular scales, $\ell\simeq 100$. 

\begin{figure}[ht]
    \centering
    \includegraphics[width=0.5\linewidth]{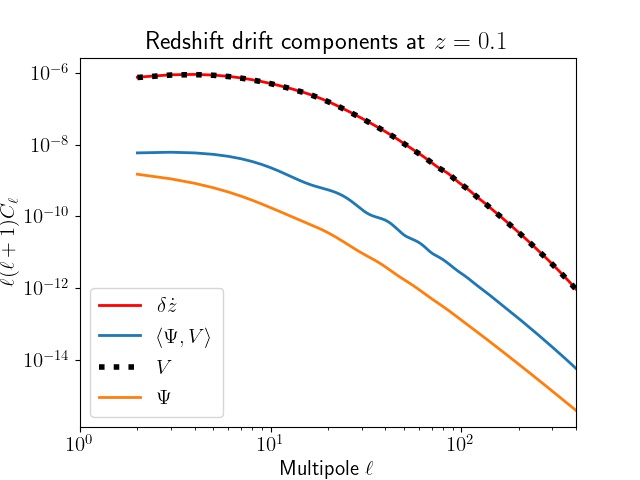}\hfil
    \includegraphics[width=0.5\linewidth]{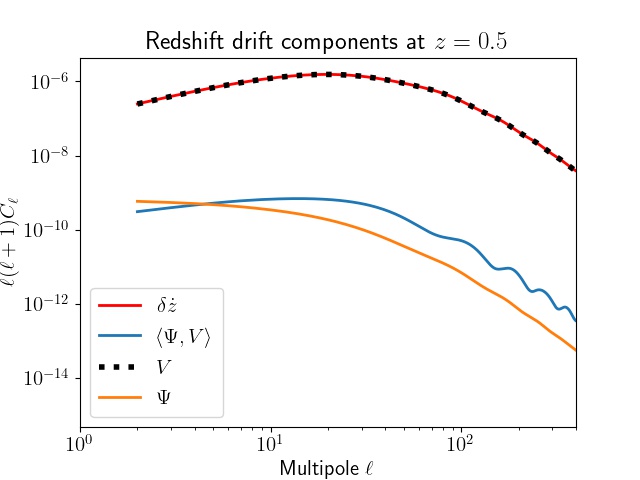}\par\medskip
    \includegraphics[width=0.5\linewidth]{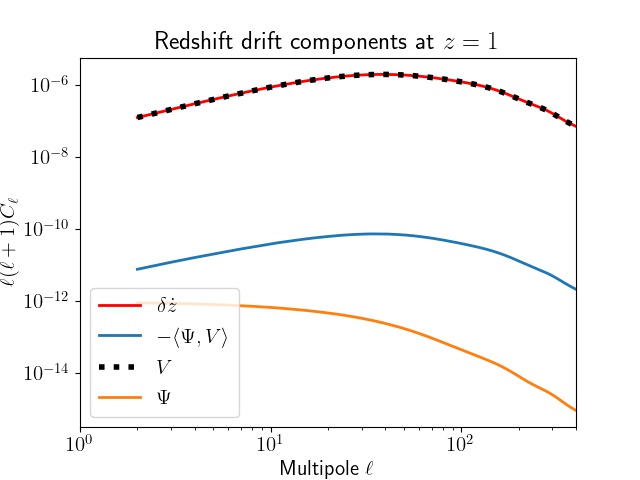}\hfil
    \includegraphics[width=0.5\linewidth]{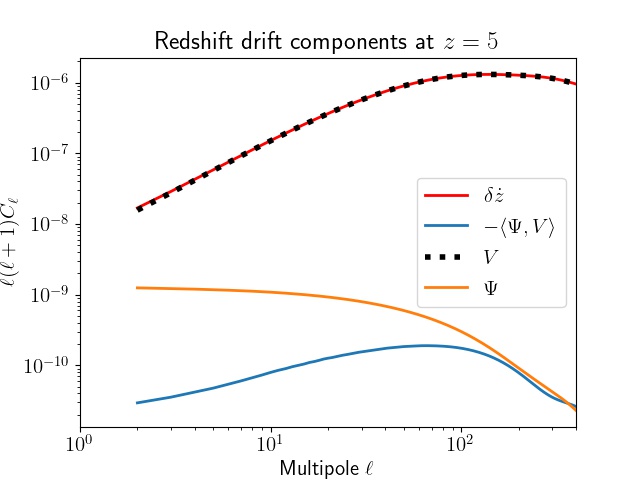}
    \caption{Redshift drift and independent components angular power spectra at different redshifts as function of $\ell$. The plot is in log-log scale and we use a Gaussian window function of width $\Delta z=0.01$. The cross correlations $\langle\Psi|V\rangle$ are positive at lower redshifts ($z=0.1, 0.5$), but turn negative at higher redshifts ($z=1,5$).}
    \label{drift_figure_1}
\end{figure}

The dominating contribution from the peculiar velocity and acceleration terms, especially at large scales and over the redshift ranges smaller than unity highlights the importance of this term in the full gauge-invariant expression for the redshift drift \eqref{eq:main_redshiftdrift}, corroborating the analytical estimates such as the one in \cite{Kim_2015} and improving on works that neglect or discard the effect of  peculiar acceleration and velocity of sources in their expression for the perturbed redshift drift \cite{Marcori_2018,Koksbang_2023}. In 
Refs.~\cite{Liske_2008} and \cite{Cooke_2019} the authors discuss, based on the use of hydrodynamical simulations, possible astrophysical sources where the peculiar acceleration and velocity can be distinguished from the background redshift drift, aiming for an increase in the detectability of the signal. Fig.~\ref{drift_figure_1} indicates that the drift perturbations are dominated by the velocity terms at all proposed observational redshift ranges. 
Nevertheless, our results show that for the redshift drift signal from several sources averaged over an angular patch of a degree or more at $z=1$ the background signal will be about 1000 times larger than the perturbations from peculiar velocities and acceleration, thus, linear perturbation theory can be trusted.
Furthermore, the contribution from the potential and cross correlated terms are at least another order of magnitude smaller at all scales.  

\begin{figure}[ht]
    \centering
    \includegraphics[width=0.5\linewidth]{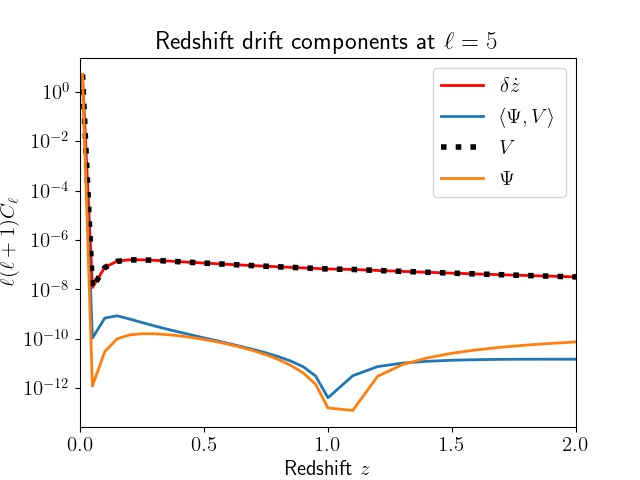}\hfil
    \includegraphics[width=0.5\linewidth]{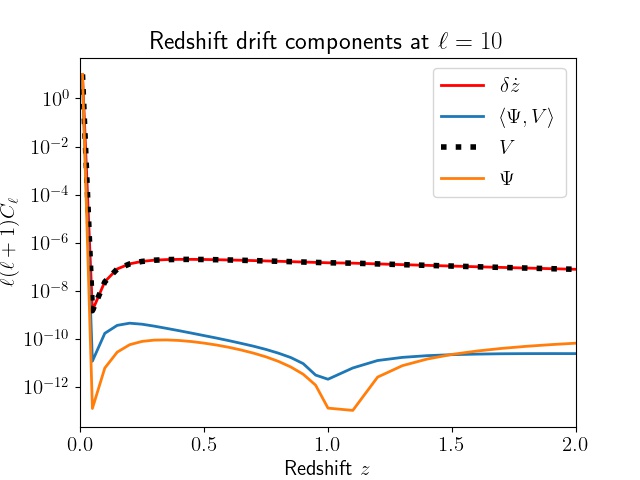}\par\medskip
    \includegraphics[width=0.5\linewidth]{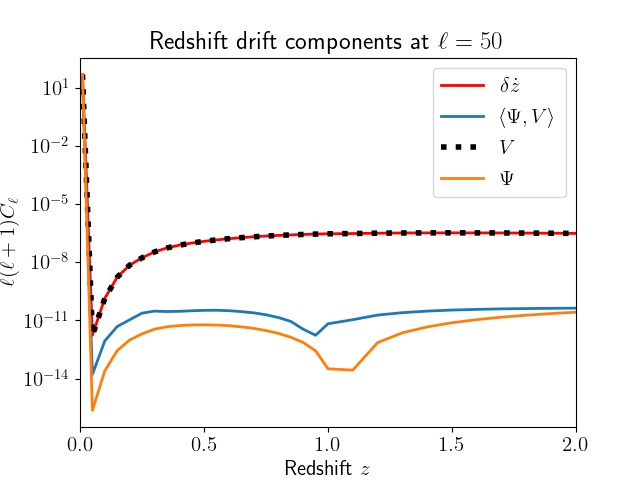}\hfil
    \includegraphics[width=0.5\linewidth]{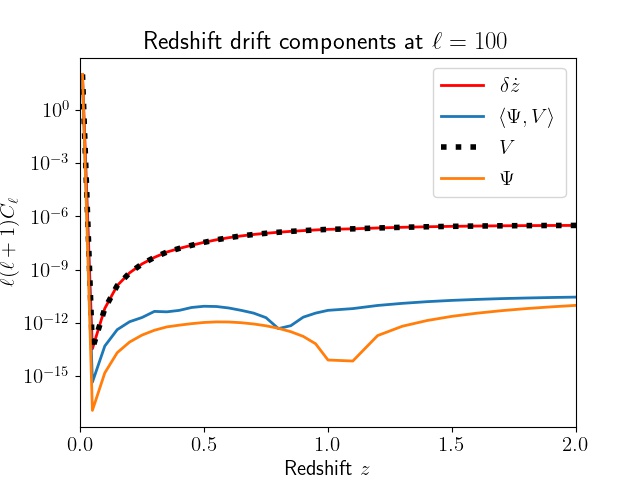}
    \caption{Redshift drift and components power spectra as a function of redshift. We chose a set of $\ell$s to characterize the scale dependence found in Fig.~\ref{drift_figure_1}.}
    \label{C_l_z}
\end{figure}

In Fig.~\ref{C_l_z} we show in detail the redshift dependence of each component. The $V$ terms keep dominating at all scales, while the potential and velocity cross-correlation terms are somewhat larger  at all scales for small redshifts. The dip in both of these contributions at around $z\approx 1$ comes from the pre-factor $\dot{\mathcal{H}}/\HH$  for $F_l^{\Psi}$, see Table~\ref{tab:legend_fig}, which changes sign around this redshift.

\begin{table}[ht]
\begin{tabular}{|l|l|l|}
\hline
$ F^{V}_\ell(z)$ & $\frac{1}{\mathcal{H}}\left(-\Dot{T}_V(k,t_s)+\frac{\mathcal{H}^2-\Dot{\mathcal{H}}}{\mathcal{H}}T_V(k,t_s)\right)j_\ell'(kr_s)$\\
$ F^{\Psi}_\ell(z)$ & $\frac{1}{\mathcal{H}}\left(\Dot{T}_\Psi(k,t_s)-\frac{\Dot{\mathcal{H}}}{\mathcal{H}}T_\Psi(k,t_s) \right)j_\ell(kr_s)$\\
$ F^{V}_\ell(z)\cdot F^{\Psi}_\ell(z)$&$\frac{1}{\mathcal{H}^2}\left(-\Dot{T}_V(k,t_s)+\frac{\mathcal{H}^2-\Dot{\mathcal{H}}}{\mathcal{H}}T_V(k,t_s)\right)\cdot\left(\Dot{T}_\Psi(k,t_s)-\frac{\Dot{\mathcal{H}}}{\mathcal{H}}T_\Psi(k,t_s) \right)j_\ell(kr_s)j'_\ell(kr_s)$\\
\hline
\end{tabular}
\caption{We label each component of the redshift drift power spectra in Fig.~\ref{drift_figure_1} and Fig.~\ref{C_l_z} according to the transfer functions inside each contribution to the total redshift drift, compare (\ref{eq:F_ell}).}
\label{tab:legend_fig}
\end{table}

We have also calculated the correlation between the redshift drift spectra at different $z$ by slightly altering \eqref{power_spectra_rd} to integrate over the kernels $F_\ell(k,z)$ at different redshifts:

\begin{equation}
\label{power_spectra_different_z}
    C^{\delta\dot{z}}_\ell(z_1,z_2) = \frac{2A}{\pi}\int \frac{dk}{k}\left(\frac{k}{k_*}\right)^{n_S-1}F_\ell(k,z_1)F_\ell(k,z_2),
\end{equation}
while the kernels are kept the same as in the original expression. 

The correlations are at most of the same order of magnitude of the power spectra measured at a single redshift signal, and in general several orders of magnitudes smaller. This small effect for non-diagonal correlations does not give any new information about the redshift dependence of the signal.
In fact, the small correlations at different redshifts give support to the observational perspectives described for instance in \cite{Cooke_2019,Kim_2015}, where the authors enforce that measurements of the drift at different redshifts, for sources with similar spectra, should improve significantly the statistics of the measurement, particularly over the peculiar acceleration of the sources. As the correlation of perturbations from  different redshifts can be safely neglected, the constraining power for the evolution of $\Dot{z}$ will improve when more redshift bins are observed.

\subsection{Cross-correlations with galaxy number counts}

We also study  the correlation between the  redshift drift \eqref{power_spectra_rd} and the theoretical relativistic number counts of galaxies at linear order. We follow the standard treatment of \cite{Bonvin_2011}, which is also manifestly gauge-invariant and therefore easily mapped to observables for checking our predictions with forthcoming surveys and simulations.

The relativistic galaxy number count fluctuation spectrum can be expressed as 

\begin{equation}
C^{N}_\ell(z_s) = \frac{2A}{\pi}\int\frac{dk}{k}(kt_O)^{n-1} \left|F^{N}_\ell(k,z_s)\right|^2,
\end{equation}
where the kernel is given by~\cite{Bonvin_2011}
\begin{align}
F^{N}_\ell(k,z_s) &=& j_\ell(kr_s)\!\left[\!T_D +\left(\!1\! +\!\frac{\dot \HH}{\HH^2} \!+
 \!\frac{2}{r_s\HH}\!\right)T_\Psi   \!+ \! T_\Phi \!+ \! \frac{1}{\HH}\dot T_\Phi \!\right] 
+ j_\ell'(kr_s)\left(\!\frac{\dot \HH}{\HH^2} \!
+\! \frac{2}{r_s\HH} \right)T_V  \!+\! \frac{k}{\HH}T_V j_\ell''(kr_s) 
   \nonumber \\  && 
  \hspace{-2cm} + \frac{1}{r_s}\int_{0}^{r_s}j_\ell(k\la)\left(2 +\frac{r_s-\la}{\la}\ell(\ell+1)\right)
   (T_\Psi + T_\Phi)d\la   + \left(\frac{\dot \HH}{\HH^2} + \frac{2}{r_s\HH}\right)   
\int_{0}^{r_s}j_\ell(k\la)(\dot T_\Psi +\dot T_\Phi)d\la,
\label{kernel_nc}
\end{align}
and the density contrast transfer function $T_D$ is related to the Bardeen potential and velocity transfer functions \eqref{eq:T_v} through Einstein's equations:

\begin{equation}
T_D= -\frac{2 a}{3\Om_m}\left(\frac{k}{\HH_0}\right)^2T_\Psi-3T_\Psi-3\frac{\HH}{k}T_V.
\end{equation}

a complete derivation of this expression can be found in e.g. \cite{Bonvin_2011,durrer_2020}. By denoting the redshift drift kernel \eqref{eq:F_ell} by $F^{\dot{z}}_\ell(k,z)$, the cross-correlation power spectrum is given by

\begin{equation}
    C^{\delta\dot{z}N}_\ell(z) \equiv \frac{A}{\pi}\int \frac{dk}{k}\left(\frac{k}{k_*}\right)^{n_S-1}\left(F^{\delta\dot{z}}_\ell(k,z)F^N_\ell(k,z)\right).
    \label{cross_corr}
\end{equation}

\begin{figure}[ht]
    \centering
    \includegraphics[scale=.90
]{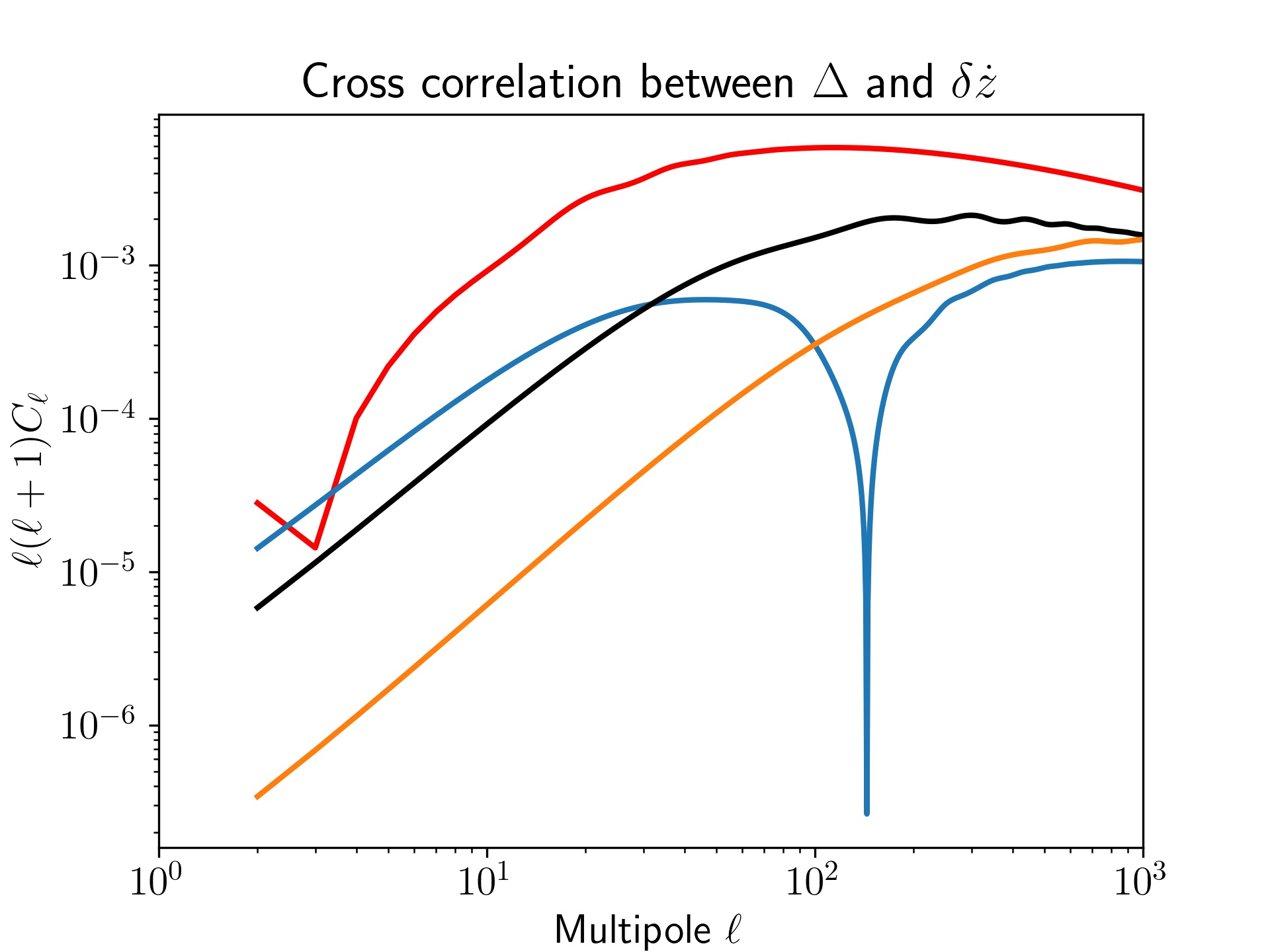}
    \caption{The cross correlation $C_\ell^{\delta\dot{z}\Delta}$ of redshift drift $\delta\dot{z}$ and relativistic number count perturbations $\Delta$ as a function of $\ell$ for different values of redshift: $z=0.1$ in red, $z=0.5$ in blue, $z=1$ in black, and $z=5$ in yellow. The spectrum is negative for $z=0.1$, and for $z=0.5$ at low $\ell$. For $z=0.5$, a sign change from  negative to positive happens around $\ell\approx 150$. All other spectra are positive.}
    \label{cross_correlation}
\end{figure}

We plot this cross correlation power spectrum for different redshifts in Fig.~\ref{cross_correlation}.

As we see in the figure, there is a similar behavior for all redshifts, with the absolute value of the amplitude increasing from large to intermediary scales and then flattening at smaller scales. For redshift values $z\lesssim 0.5$ however, the signals are anticorrelated for some ranges of $\ell$: for $z=0.1$ the signal is strictly anticorrelated, while for $z=0.5$ the signal is anticorrelated for scales $\ell\lesssim150 $, where it changes sign.
For higher values of $\ell$ the curves reach a maximum, respectively a minimum, at $\ell_{\text{max}}(z)$ and have a slow decrease/increase up to highly nonlinear scales $\ell\approx 1000$, where we see a  plateau between intermediate and small scales.
This result is interesting because, as it has been argued in \cite{Liske_2008}, that peculiar acceleration and velocities are the perturbations which are most difficult to distinguish from the cosmic background signal in future observational probes of redshift drift. Therefore, an observational probe that is correlated with the peculiar acceleration such as the number counts could provide a way to distinguish these effects from the true background signal.
We notice that the maximum of the amplitude increases with redshift, such that for redshifts $z>0.5$ the curves reach it at $\ell_{\text{max}}>100$. This behavior, and the anticorrelation for smaller redshifts, indicates that the $z\lesssim 0.5$ redshift range has a characteristic cross-correlation spectra which could possibly help to strenghten the detection of the $\delta \dot{z}$ signal through the correlation. The measured drift from sources at different redshifts in this redshift range can be correlated with the galaxy number count, and the change of sign observed in Fig. \ref{cross_correlation} for the same scale and different redshifts in this range could be a characteristic feature of the redshift drift signal, distinguishing it from the $\delta\dot{z}$ self-correlation power spectrum, which is strictly positive, see Fig. \ref{power_spectra_l_500}.

 This redshift range has also been cited as promising in detecting the redshift drift precisely due to its small dependence on other cosmological probes \cite{Kim_2015}. Furthermore, the correlation amplitude is about three orders of magnitude higher than the amplitude of  $\Dot{z}$, as one can see by comparing Fig.~\ref{cross_correlation} with Fig.~\ref{power_spectra_l_500}.

Having different redshift ranges where the correlation is significant in the $\Lambda$-dominated and matter-dominated eras as well as the change from anti-correlation to correlation in the transition epoch can provide additional constraints for the dark energy equation of state, as explored at the background level in~\cite{Kim_2015}. The curves in Fig.~\ref{cross_correlation} show that small scales provide the highest correlations for redshifts $z\gtrsim0.5$, while at larger scales number count perturbations are more strongly correlated with fluctuations of the drift at lower redshifts, inside the $\Lambda$-dominated era. Furthermore, there is a change from anti-correlation to correlation as the redshift increases. This behavior may provide insight into the redshift dependence of the drift, and particularly how possible tracers of the effect should be chosen, as the sign and amplitude of the cross-correlation change, especially for smaller redshifts.

In all the cases discussed, the cross-correlation signal has a higher amplitude than the $\delta\dot{z}$ power spectra alone. On average this difference is about three orders of magnitude for all redshifts. This correlation may be useful to both strengthen the significance in the detection of the signal and to constrain the cosmological model through the redshift drift signal, such as in the case for the ISW effect \cite{Nishizawa_2014}.

The question whether this correlation signal can truly be measured in one of the proposed surveys, and if yes with which significance, however, is a more difficult question. This  depends on the number density of observed sources and on the accuracy of the measurement of their redshift drift. Nevertheless, correlating the signal with galaxy number counts is boosting it by about a factor $\sqrt{ C^{\delta\dot{z}N}/ C^{\delta\dot{z}}}\sim 100$. Furthermore, the cross-correlation of number counts and  redshift drift does not suffer from shot noise and can provide a significant signal already with  a lower number density of galaxies.
A detailed calculation of the signal to noise of such a measurement for a given experiment goes beyond the present study.
\section{Discussion and conclusion}
\label{sec:con}

In this paper we have derived the fully gauge-invariant perturbation of the cosmological redshift drift and we have numerically calculated its power spectra over a wide range of redshifts and scales within linear perturbation theory. Our derivation is general and includes all terms up to first order, expanding and generalizing previous results that relied on source or observer dependent assumptions.

Our main results for the perturbed cosmic redshift drift, expressions \eqref{eq:main_redshiftdrift} and \eqref{eq:reddrift_final}, are manifestly gauge-invariant. They generalize previous expressions for the perturbed redshift drift, as found in e.g.~\cite{Marcori_2018,Liske_2008}, and make it straightforward to derive constraints on this quantity from survey and simulation data. We expect the theoretical results to translate easily to numerical codes and simulations, as has been done in this paper through the use of {\sc class}.

We find that within linear perturbation theory redshift drift fluctuations are small, typically of order $10^{-3}$ or smaller. Of course, a single redshift drift measurement might be strongly affected by peculiar acceleration of the source. However, averaging the measured redshift drift at fixed redshift over many widely separated directions, as proposed e.g. in~\cite{Cooke_2019}, will yield a good measure of the background redshift drift, as in this situation the fluctuations of this mean are well-described by linear perturbation theory.
Nevertheless, planned measurements, e.g. with SKA2, will need $N\simeq 10^7$ galaxies in each redshift bin to reach their goal~\cite{Klockner:2015rqa}. They assume that the signals from the galaxies are uncorrelated so that the total error is given by the shot noise term, $1/\sqrt{N}$. Our findings show that they have to add an effect from the correlations which is of a similar order of magnitude. 
More precisely, we obtain a variance from the correlations of
\be
\si_{\de z}^2(z) =\frac{1}{4\pi}\sum_\ell(2\ell+1)C_\ell(z) \,.
\ee

\begin{table}[!ht]
\begin{center}\begin{tabular}{|c|c|c|}
\hline
$z$ & $\si_{\de z}$ & $1/\sqrt{N(z)} $\\
\hline
0.1 &$7\times10^{-4}$  & ~ $1.8\times 10^{-4}$~  \\
0.5 & $1\times 10^{-3}$  & ~ $3.2\times 10^{-4}$~  \\
1 &$1.1\times 10^{-3}$ & ~ $1.8\times 10^{-3}$~  \\
5 &$9.5\times 10^{-4}$   & ?  \\
\hline
\end{tabular}
\end{center}
\caption{ \label{t:sig}The expected error from correlations for a full sky experiment of the redshift drift at $z=0.1,~0.5,~1$ and $5$. The expected numbers of galaxies for an SKA2 survey are taken from~\cite{Klockner:2015rqa} (Fig.~2 lower panels). SKA2 only plans to measure redshift drift out to $z=1$. }
\end{table}
In Table~\ref{t:sig} we give the values of $\si_{\de z}(z)$ for some redshifts.
Clearly, the correlation contribution to the error is comparable with shot noise and has to be added. For the E-ELT the situation is different. They plan to measure the redhift drift much more precisely, i.e. with much smaller errors per measurement, but only for order ten lines of sight per redshift bin~\cite{Cristiani:2023iwi}. The shot noise contribution is then much larger, such that the correlation error is irrelevant.

The dominant contribution to first order perturbations comes from the peculiar velocity and acceleration of sources: this term is more than two orders of magnitude larger than the contribution  from the gravitational potential  for most redshifts, as it is seen in Fig.~\ref{drift_figure_1}.  Furthermore, the acceleration and potential terms are very weakly correlated. While the relevance of the peculiar acceleration and velocity of sources as perturbative effects on the cosmic redshift drift have been previously studied in the literature~\cite{Liske_2008,Quercellini_2012}, our results show for the first time the dominance of this term in comparison to other contributions to the perturbed redshift drift signal; moreover previous results have all  relied on gauge-dependent derivations of the perturbed redshift drift.
Here, the theoretical power spectra for the redshift drift perturbations were calculated for the first time and its numerical evaluation  is presented in Fig.~\ref{power_spectra_l_500}. The scale dependence of the effect has not been previously studied and the power spectrum prediction quantifies this, paving the way for further observational probes of the effect. We have also studied the evolution of the signal with redshift in Fig.~\ref{redshift_drift_z}.

Finally, we have also investigated the cross-correlation with galaxy number counts, and find significant correlations at all scales, which increases with $\ell$ and peaks at intermediary scales for smaller redshifts and small scales for high redshifts. In particular, for smaller redshifts $z\lesssim 0.5$, there is a change of sign from anti-correlated to correlated as the redshift increases. This correlation might help in the detection of the cosmic redshift drift fluctuations by strengthening the power spectra signal. Furthermore, the scale dependence of the cross-correlation power spectrum could point to possible astrophysical probes of the effect, such as explored in \cite{Cooke_2019}.\\

\section*{Acknowledgement}
We thank Enea Di Dio for help with the {\sc class}gal code and useful discussions.
PB acknowledges financial support from Coordenação de Aperfeiçoamento de Pessoal de Nível Superior (CAPES) and Fundação de Apoio à Pesquisa do Espírito Santo (FAPES) for his PhD fellowship and is very thankful for the hospitality of the  Université de Genève, where most of this project was worked out. RD and DS acknowledge support by the Swiss National Science Foundation under grant No 200020\underline{~}182044.
\vspace{2cm}
\appendix

\section{Appendix: Gauge-invariant Perturbation Variables}
\label{ap:perturbationvariables}
This appendix summarises some of the relations between the perturbation variables used in Section~\ref{sec:deltaz_perturbedFLRW} needed to arrive at (\ref{eq:main_redshiftdrift}). For a more detailed account we refer to \cite{durrer_2020}, here, we restate the relations relevant for this work. We first introduce scalar potentials for the velocity perturbation $v^i$, the perturbations of the 0i-elements of the metric $B^i$, the 4-acceleration $a^\mu$, and the shear tensor $\sigma_{\mu\nu}$:
\begin{align}
    v^i&=-\partial^i v\quad,\\
    B^i&=-\partial^i B\quad,\\
    a^i &= -\partial^i \mathcal{A}\quad,\quad a^0=0\quad\text{with}\quad \\
    \mathcal{A}& =\frac{1}{a^2}\left(\dot{B}-\dot{v}+A+\mathcal{H}(B-v)\right) =\frac{1}{a^2}\left(-\dot V -\HH V +\Psi \right) \ \,,\\
    \sigma_{ij} &= a\left(\partial_i \partial_j -\frac{1}{3}\delta_{ij}\Delta\right)\sigma\quad,\quad \sigma_{00}=0=\sigma_{0i}=\sigma_{i0}\quad.
\end{align}
The final result (\ref{eq:main_redshiftdrift}) is expressed in terms of the gauge-invariant velocity perturbation $V$, the Bardeen potentials $\Phi$ and $\Psi$, and the spatial curvature perturbation $\mathcal{R}$, which are related to the above mentioned perturbation variables via:
\begin{align}
    V &=v-\dot{H}_T\quad,\\ 
    \Psi &= A-\mathcal{H}\sigma_t-\dot{\sigma}_t\quad,\\
    \Phi &= -\mathcal{R}+\mathcal{H}\sigma_t\quad.
\end{align}
The following relations between the perturbation variables are useful:
\begin{align}
\mathcal{R} &= H_L-\frac{1}{3}\Delta H_T  \quad,\\
    \sigma &= -V= \dot{H}_T-v\quad,\\
    \sigma_t &= \dot{H}_T-B\quad,\\
    V &= v-B-\sigma_t\quad.
\end{align}
Note that while $\si= -V$ is gauge invariant, $\mathcal{R}$ and $\si_t$ are not. The denote the spatial curvature and the shear on the $t=$ constant hypersurface and therefore depend on the chosen time-slicing.

\bibliography{paper_v2}

\begin{thebibliography}{10}

\bibitem{Sandage:1962}
A.~Sandage.
\newblock The change of redshift and apparent luminosity of galaxies due to the
  deceleration of selected expanding universes.
\newblock {\em Astrophys. J.}, 136, 1962.

\bibitem{McVittie:1962}
G.~McVittie.
\newblock Appendix to the change of redshift and apparent luminosity of
  galaxies due to the deceleration of selected expanding universes.
\newblock {\em Astrophys. J.}, 136, 1962.

\bibitem{Liske:2008ph}
J.~Liske et~al.
\newblock {Cosmic dynamics in the era of Extremely Large Telescopes}.
\newblock {\em Mon. Not. Roy. Astron. Soc.}, 386:1192--1218, 2008.

\bibitem{Rocha:2022gog}
B.~A.~R. Rocha and C.~J. A.~P. Martins.
\newblock {Redshift drift cosmography with ELT and SKAO measurements}.
\newblock {\em Mon. Not. Roy. Astron. Soc.}, 518(2):2853--2869, 2022.

\bibitem{Koksbang_2020_2}
S~M Koksbang.
\newblock {Observations in statistically homogeneous, locally inhomogeneous
  cosmological toy models without FLRW backgrounds}.
\newblock {\em Monthly Notices of the Royal Astronomical Society: Letters},
  498(1):L135--L139, 09 2020.

\bibitem{Koksbang_2021_prl}
S.{\hspace{0.167em} }M. Koksbang.
\newblock Searching for signals of inhomogeneity using multiple probes of the
  cosmic expansion rate h(z).
\newblock {\em Physical Review Letters}, 126(23), jun 2021.

\bibitem{Heinesen_2021}
Asta Heinesen.
\newblock Redshift drift as a model independent probe of dark energy.
\newblock {\em Physical Review D}, 103(8), apr 2021.

\bibitem{Marcori_2018}
Oton~H. Marcori, Cyril Pitrou, Jean-Philippe Uzan, and Thiago~S. Pereira.
\newblock Direction and redshift drifts for general observers and their
  applications in cosmology.
\newblock {\em Physical Review D}, 98(2), jul 2018.

\bibitem{Heinesen:2020pms}
Asta Heinesen.
\newblock {Multipole decomposition of redshift drift -- model independent
  mapping of the expansion history of the Universe}.
\newblock {\em Phys. Rev. D}, 103(2):023537, 2021.

\bibitem{ellis_maartens_maccallum_2012}
George F.~R. Ellis, Roy Maartens, and Malcolm A.~H. MacCallum.
\newblock {\em Relativistic Cosmology}.
\newblock Cambridge University Press, 2012.

\bibitem{durrer_2020}
Ruth Durrer.
\newblock {\em The Cosmic Microwave Background}.
\newblock Cambridge University Press, 2 edition, 2020.

\bibitem{Planck:2018jri}
Y.~Akrami et~al.
\newblock {Planck 2018 results. X. Constraints on inflation}.
\newblock {\em Astron. Astrophys.}, 641:A10, 2020.

\bibitem{Inoue:2019qvy}
Takuya Inoue, Eiichiro Komatsu, Wako Aoki, Takeshi Chiba, Toru Misawa, and
  Tomonori Usuda.
\newblock {The effect of our local motion on the Sandage\textendash{}Loeb test
  of the cosmic expansion}.
\newblock {\em Publ. Astron. Soc. Jap.}, 72(1):131, 2020.

\bibitem{Bonvin_2011}
Camille Bonvin and Ruth Durrer.
\newblock What galaxy surveys really measure.
\newblock {\em Physical Review D}, 84(6), sep 2011.

\bibitem{Planck_18}
Planck Collaboration.
\newblock Planck 2018 results. vi. cosmological parameters.
\newblock {\em Astronomy \& Astrophysics}, 641:A6, September 2020.

\bibitem{dodelson2003}
Scott Dodelson.
\newblock {\em {Modern Cosmology}}.
\newblock Academic Press, Elsevier Science, 2003.

\bibitem{CLASS2}
Diego Blas, Julien Lesgourgues, and Thomas Tram.
\newblock The cosmic linear anisotropy solving system ({CLASS}). part {II}:
  Approximation schemes.
\newblock {\em Journal of Cosmology and Astroparticle Physics},
  2011(07):034--034, jul 2011.

\bibitem{Eisenstein_1998}
Daniel~J. Eisenstein and Wayne Hu.
\newblock Baryonic features in the matter transfer function.
\newblock {\em The Astrophysical Journal}, 496(2):605--614, apr 1998.

\bibitem{Orjuela22}
J.~Bayron Orjuela-Quintana, Savvas Nesseris, and Wilmar Cardona.
\newblock Using machine learning to compress the matter transfer function
  $t(k)$.
\newblock {\em Physical Review D}, 107(8), apr 2023.

\bibitem{DiDio:2013bqa}
Enea Di~Dio, Francesco Montanari, Julien Lesgourgues, and Ruth Durrer.
\newblock {The CLASSgal code for Relativistic Cosmological Large Scale
  Structure}.
\newblock {\em JCAP}, 11:044, 2013.

\bibitem{Kim_2015}
Alex~G. Kim, Eric~V. Linder, Jerry Edelstein, and David Erskine.
\newblock Giving cosmic redshift drift a whirl.
\newblock {\em Astroparticle Physics}, 62:195--205, mar 2015.

\bibitem{Koksbang_2023}
Sofie~Marie Koksbang.
\newblock Redshift drift in a universe with structure. {II}. light rays
  propagated through a newtonian n-body simulation.
\newblock {\em Physical Review D}, 107(6), mar 2023.

\bibitem{Liske_2008}
J.~Liske, A.~Grazian, E.~Vanzella, M.~Dessauges, M.~Viel, L.~Pasquini,
  M.~Haehnelt, S.~Cristiani, F.~Pepe, G.~Avila, P.~Bonifacio, F.~Bouchy,
  H.~Dekker, B.~Delabre, S.~D{\textquotesingle}Odorico,
  V.~D{\textquotesingle}Odorico, S.~Levshakov, C.~Lovis, M.~Mayor, P.~Molaro,
  L.~Moscardini, M.~T. Murphy, D.~Queloz, P.~Shaver, S.~Udry, T.~Wiklind, and
  S.~Zucker.
\newblock Cosmic dynamics in the era of extremely large telescopes.
\newblock {\em Monthly Notices of the Royal Astronomical Society},
  386(3):1192--1218, may 2008.

\bibitem{Cooke_2019}
Ryan Cooke.
\newblock The {ACCELERATION} programme: I. cosmology with the redshift drift.
\newblock {\em Monthly Notices of the Royal Astronomical Society},
  492(2):2044--2057, dec 2019.

\bibitem{Nishizawa_2014}
A.~J. Nishizawa.
\newblock The integrated sachs-wolfe effect and the rees-sciama effect.
\newblock {\em Progress of Theoretical and Experimental Physics},
  2014(6):6B110--0, jun 2014.

\bibitem{Klockner:2015rqa}
Hans-Rainer Kl\"ockner, Danail Obreschkow, Carlos Martins, Alvise Raccanelli,
  David Champion, Alan~L. Roy, Andrei Lobanov, Jan Wagner, and Reinhard Keller.
\newblock {Real time cosmology - A direct measure of the expansion rate of the
  Universe with the SKA}.
\newblock {\em PoS}, AASKA14:027, 2015.

\bibitem{Quercellini_2012}
Claudia~Quercellini et~al.
\newblock Real-time cosmology.
\newblock {\em Physics Reports}, 521(3):95--134, dec 2012.

\end{thebibliography}
\bibliographystyle{unsrt}
\end{document}